# Towards Semi-automatic Detection and Localization of Indoor Accessibility Issues using Mobile Depth Scanning and Computer Vision

Xia Su, Allen School of Computer Science, University of Washington

Kaiming Cheng, Allen School of Computer Science, University of Washington

Han Zhang, Allen School of Computer Science, University of Washington

Jaewook Lee, Allen School of Computer Science, University of Washington

Yueqian Zhang, Allen School of Computer Science, University of Washington

Jon E. Froehlich, Allen School of Computer Science, University of Washington

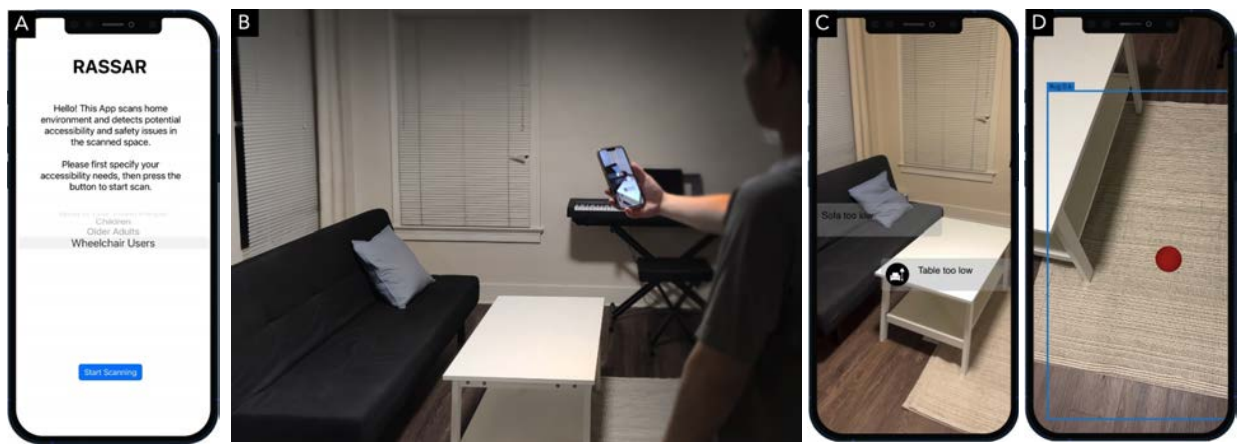

**Figure 1.** In this workshop paper, we present a proof-of-concept prototype called *RASSAR (Room Accessibility and Safety Scanning in Augmented Reality)* for *identifying*, *categorizing*, and *localizing* indoor accessibility and safety issues using LiDAR + camera data, real-time machine learning, and AR. With RASSAR, the user first selects the target community (a), scans the indoor space with their mobile phone (b), and receives real-time feedback about problematic items, dimensions, and positions. The feedback is anchored on the object's location in real-world coordinate space (c and d).

# Abstract

To help improve the safety and accessibility of indoor spaces, researchers and health professionals have created assessment instruments that enable homeowners and trained experts to audit and improve homes.  With advances in computer vision, augmented reality (AR), and mobile sensors, new approaches are now possible. We introduce *RASSAR (Room Accessibility and Safety Scanning in Augmented Reality)*, a new proof-of-concept prototype for semi-automatically *identifying*, *categorizing*, and *localizing* indoor accessibility and safety issues using LiDAR + camera data, machine learning, and AR. We present an overview of the current RASSAR prototype and a preliminary evaluation in a single home.

# Introduction

Safe, accessible housing is a fundamental human right [47]. Indeed, the UN's *Convention on the Rights of Persons with Disabilities* states that governments should identify and eliminate accessibility barriers not just in public facilities and transportation but in schools, workplaces, and homes [51]. To help improve the safety and accessibility of domestic spaces, researchers and health professionals have created pre-formatted checklists (assessment instruments) that enable homeowners and trained professionals to audit and improve indoor spaces [13,19,20,22,37,40,42]. Often, these instruments are designed for particular target groups, such as older adults [13,19,36,37], children [27,40], and/or people with specific disabilities [10,22]. Some instruments, including *SAFER HOME* [17] and the *WeHSA* [11], were created for professionally trained occupational therapists to conduct in conjunction with home visits. Others have explored self-assessment tools such as the *Home Safety Self-Assessment Tool* (HSSAT) [19,42] and *Remote Home Safety Protocol* [37], which reduce cost, eliminate the need for professional intervention, and improve homeowner education. HSSAT, for example, contains a safety and accessibility checklist for nine areas of the home, including entrances, kitchens, bathrooms, and bedrooms, and includes issues such as a lack of wheelchair ramp, uneven flooring, cluttered areas, presence of throw rugs, electric cords across the floor, and inaccessible light switches.

With advances in computer vision, augmented reality (AR), and mobile sensors, new approaches for assessing indoor accessibility and safety are now possible. In this paper, we introduce *RASSAR (Room Accessibility and Safety Scanning in Augmented Reality)*, a novel smartphone-based prototype for semi-automatically *identifying*, *categorizing*, and *localizing* indoor accessibility and safety issues using LiDAR + camera data, real-time machine learning, and AR. With RASSAR, the user holds out their phone and slowly scans a space—the tool constructs a real-time parametric model of the 3D scene, attempts to identify and classify known accessibility and safety issues, and visualizes potential problems overlaid in AR—see Figure 1. We envision RASSAR as a tool to help build and validate new construction, for homeowners planning renovations or updating their homes due to a life change (*e.g.*, illness, birth), or for rental services like Airbnb to help vet and validate access and safety.

As initial work, RASSAR currently detects 18 objects and examines issues such as *object dimension* (*e.g.*, table too low), *position* (*e.g.*, cabinet too high), and *existence* (*e.g.*, throw rug). These supported features are informed by ADA standards for accessible design [43,52] and the health and safety literature [33]. We present an overview of the RASSAR prototype, its custom-trained machine learning model, and an initial evaluation in a single home. Our work contributes a new approach to assess indoor access/safety issues using emerging technology.

## Background and Related Work

RASSAR automatically constructs 3D models of the indoor environment using the iPhone Pro's built-in LiDAR scanner and classifies potential access and safety problems using LiDAR and RGB camera data. We cover work related to both below.

In architecture and construction fields, LiDAR is increasingly used to semi-automatically generate 2D blueprints and create interactive 3D models of buildings, often to track construction progress or examine a building post-construction (*e.g.*, compare "as-designed" blueprints to "as-built" results) [23,34,45]. Here, the primary focus is on generating accurate 2D/3D blueprints of rooms, including room geometry, wall locations and orientations, and the position and size of windows and doors. Interior objects like furniture are deemed "visual clutter," which occlude laser scans and negatively impact building model generation [18,23]. In our research, we are interested in both the geometry and layout of space as well as the furniture and other objects within it and how each may pose an accessibility or safety issue.

In computer vision, combining LiDAR and camera data for scene reconstruction and object identification is an active area of research [29,35], particularly for autonomous vehicles [28,49]. A key challenge is how to leverage both RGB and depth sensor streams to improve detection results [3]. Others have examined purely 3D-detection approaches that classify objects directly from point cloud information [2,8]. In our case, we use both LiDAR and camera data.

Most relevant to our work are applications of sensing and computer vision to support indoor accessibility and safety. In this space, most work focuses on real-time navigation for blind or low-vision users, such as navigating hallways [25] or stairs [50], supporting pedestrian position estimations [24], or general wayfinding through localization and landmark detection [12,15,48]. While nascent, this work is already emerging in commercial tools. In May 2022, Apple introduced *Door Detection*, which fuses real-time LiDAR+camera data with on-device machine learning to help blind and low-vision users locate doors, understand distances to doorways, and describe door attributes—including open/closed state [53].

While there is a large body of active research in automated approaches to audit and document the accessibility of outdoor pedestrian environments [1,21,26,32,38,39,46], surprisingly few works focus specifically on indoor accessibility/safety issues. Notable exceptions include Balado *et al.*'s research on automatically detecting building entrance stairs from 3D point clouds [5,6] and Ayala-Alfaro's recent work in using mobile indoor robots to scan and classify objects that may impede access for people with mobility disabilities [4]. We could find no prior work that attempts to scan, identify, and localize both accessibility and safety issues using computer vision, AR, and machine learning—which is our focus.

## The RASSAR System

RASSAR identifies, categorizes, and localizes indoor accessibility and safety issues in real-time using LiDAR + RGB camera data, computer vision, and AR (Figure 2). Specifically, we employ Apple's *RoomPlan API*[1] to create a 3D room floor plan with furniture and dimension information along with a custom-trained deep learning model (*YOLOv4* [9]) for safety/accessibility object detection using the RGB camera stream. We envision RASSAR as either a self-assessment tool for homeowners and caregivers or as a complementary tool for occupational therapists. We are currently targeting four communities of interest: children caregivers, wheelchair users, older adults, and people who are blind or low vision.

Table 1 below and Table 7 (in the Appendix) provides an overview of the 18 currently supported access/safety issues in RASSAR across three high-level categories: inaccessible or unsafe **object dimensions** (*e.g.*, door too narrow, table too low), **object positions** (*e.g.*, lightswitch position too high), and **object existence** (*e.g.*, presence of a throw rug or medication).

---

[1] https://developer.apple.com/augmented-reality/roomplan/

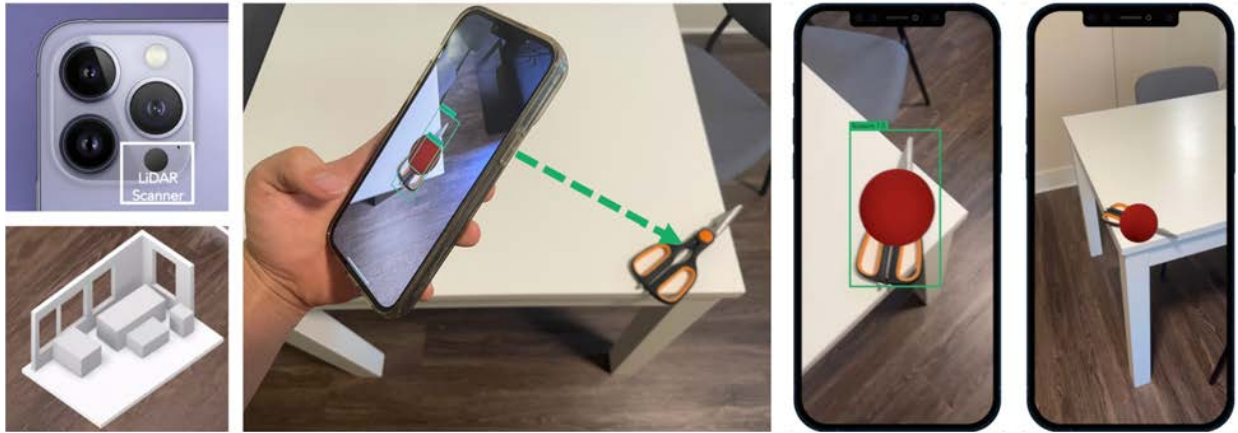

**Figure 2.** An overview of the RASSAR prototype, which uses the iPhone Pro's built-in LiDAR and RGB cameras to build a real-time model of indoor space and a custom-trained machine learning model (YOLOv4) to identify potential safety and accessibility issues. In this case, RASSAR has detected scissors. The green bounding box is from Yolo, which shows the detection confidence. We also triangulate the location of the object and insert a red sphere displayed in AR.

The RASSAR prototype is composed of three primary components:

**1. Specifying access/safety issues.** To support a variety of access/safety issues, we developed a custom input specification format in JSON. The file specifies what objects to detect (*e.g.*, table, door, knife), what problems to analyze (*e.g.*, table *height*, door *width*, sharp object *existence*), and who is potentially impacted (*e.g.*, wheelchair user, older adult). The full JSON description is in Table 1 along with two examples in Figure 3.

**Table 1.** The JSON elements and their values in our input specification file.

| JSON Element | Description | Values |
| --- | --- | --- |
| Object of interest | The object to automatically identify. We currently support 18 objects. | *Cabinet, chair, counter, door, door handle, electric socket, grab bar, knives, knob, light switch, medication, rug, sink, scissors, smoke alarm, sofa, table, toilet* |
| Community | The impacted target group(s) | *Children, older adults, blind or low vision, wheelchair user*. Can select one or more. |
| Dependency | Relational dependence (if any) | Can specify any *Object of interest*. |
| Dimension | Specifies dimensional requirements and conditional logic (*e.g.*, a door must be a minimum of 32" wide) | Comparator operators include <, ==, and >. We also support a between range check. |
| Existence | Does the object pose a problem by its sheer existence at the location? | Can be *null*, *true*, or *false*. If null, then this *Object of interest* is not about existence but about *dimension* or *position*. |
| Description | A human-readable description of the problem, which is drawn from existing ADA standards and assessment checklists | Plain text |

```
"Door-Opening":{                           "Knives":{
  "Radius":{                                 "Radius":{
    "Community":["Wheelchair User"],           "Community":["Children"],
    "Dependency": ["Door"]                     "Dependency": ["Table", "Sofa",
    "Dimension":{                               "Counter","Floor", "Bed", "Chair"],
      "Comparison":                            "Dimension":{
        ">",                                     "Comparison": null,
        "Value":[32]                             "Value": null
    },                                         },
    "Existence": null,                         "Existence": false,
     "Description": "According to ADA          "Description": "For safety, no knives
     compliance, door openings shall           should be present on the reachable
     a clear width of 32 inches   minimum."    surfaces."
  }                                            }
}                                            }
```

**Figure 3.** Two example JSON objects: (a) a doorway opening with a required width specification and (b) a check for the existence of a knife.

**2. Detecting and locating access and safety issues.** RASSAR reads in the JSON specification to initialize its problem detection engine, which relies on two core techniques: first, we employ Apple's *RoomPlan API* to build up a parametric 3D model of indoor space and use its robust built-in object recognition[2] to identify doors, walls, and windows as well as objects such as chairs, sofas, tables. For each element, we receive a 3D bounding box representing dimensional information (width, length, height) as well as a confidence score for the object classification. Currently, Apple's built-in object recognizer is limited to 16 objects: *bathtub, bed, chair, dishwasher, fireplace, oven, refrigerator, sink, sofa, stairs, storage, stove, table, television, toilet,* and *washer/dryer*.

Thus, for our second technique, we use a state-of-the-art computer vision (CV) algorithm called YOLOv4 [9]. We chose the latest YOLOv4-tiny detector as the baseline model and trained on nine additional access/safety objects using more than 2,500 images found via *Bing Image Search* and Krasin *et al.'s Open Images* [54]—see Table 2. After obtaining the weights from training, we transferred them to the AR-supported Apple *CoreML*[3] model, which is integrated into RASSAR. While YOLO does not provide 3D dimension or localization information, we *localize* recognized objects using Apple's raycast function in RealityKit[4]. Specifically, we take the center of the identified object's 2D bounding box in each video frame's image from the RGB camera and use raycasting to localize the pixel in 3D space. To reduce noise, we take the average raycast value over multiple frames (using a sliding window of *N=5* frames).

---

[2] Specifically, we use the RoomPlan's CapturedRoom object, which includes an array of doors, openings, walls, windows as well as an inferred objects array.
[3] https://developer.apple.com/documentation/coreml
[4] https://developer.apple.com/documentation/realitykit/arview

**Table 2.** The custom YOLO training set for additional access/safety objects beyond those recognized by the Apple RoomPlan API. For ground truth (GT) labeling, we used *Label Studio* (https://labelstud.io/). For each image, we drew a 2D bounding box and provided a GT label. Some images contained multiple issues.

| Object | GT Annotation | Image Count |
|---|---|---|
| Door handle | 530 | 370 |
| Electric socket | 370 | 181 |
| Light switch | 299 | 138 |
| Grab bar | 503 | 395 |
| Scissors | 270 | 226 |
| Knife | 622 | 451 |
| Medication | 688 | 325 |
| Rug | 470 | 377 |
| Smoke alarm | 191 | 176 |
| **Total** | **3,943** | **2,533** |

**3. Visualizing identified issues.** Finally, to visualize the identified issues, RASSAR currently draws a small, red sphere in AR localized in 3D space (on the object itself) and also provides a 2D pop-up that describes the problem and allows the user to confirm or disagree with the assessment. These visualizations are preliminary. We plan on providing more detail via interactive pop-ups and a full *post hoc* report when the scan is completed.

## Preliminary Evaluation

As an initial exploration of technical feasibility and performance, we conducted a controlled evaluation of RASSAR in a one-bedroom apartment. We embedded 21 access/safety problems and 10 "non-issues" in three rooms—the entrance, living room, and kitchen—see Figures 4 and 5. We then examined three factors that may impact RASSAR performance: *lighting condition*, *scanning speed*, and *room tidiness* (Table 3). Given the number of factors, it was not feasible to exhaustively test all combinations. Instead, we investigated selected combinations. For each combination, we conducted five full scans. In total, a single researcher set up and conducted tests for six combinations of factors for a total of 30 scans (Table 6). Each scan included all 21 access/safety problems and all 10 "non-issues".

For our performance measures,, we use standard metrics including *true positives*: successfully detecting an issue; *true negatives:* successfully avoiding a non-issue; *false positives*: misclassifying an issue or identifying a non-issue as a problem; and *false negatives*: not identifying an actual problem—see Table 4 for more details. As is common in machine learning research, we combine these individual metrics into aggregate measures *precision*, *recall*, $F_1$ *score*, and *accuracy* (Table 5).

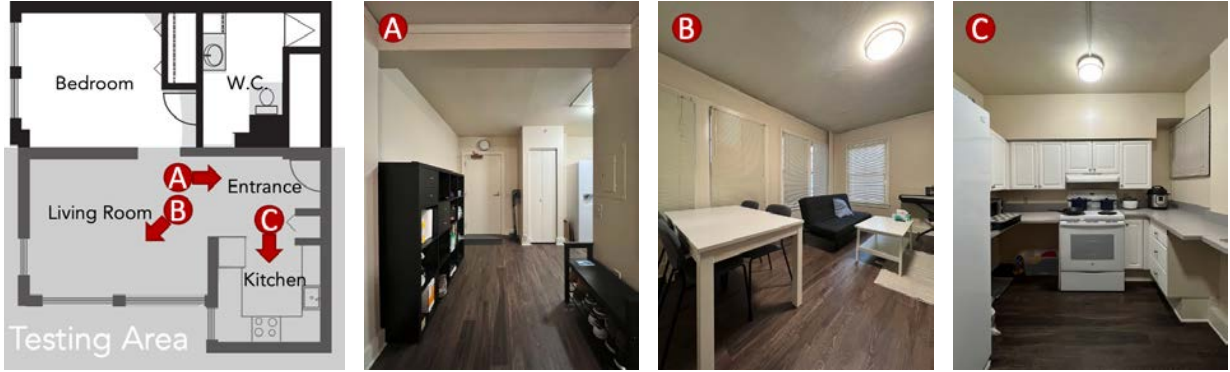

**Figure 4.** We performed an initial evaluation of RASSAR in a single apartment across three rooms: (a) the apartment entrance, (b) living room, and (c) kitchen. The red arrows in the blueprint (left image) indicate camera capture position. Note: these images are not from the RASSAR tool but, instead, are taken to illustrate the study setup.

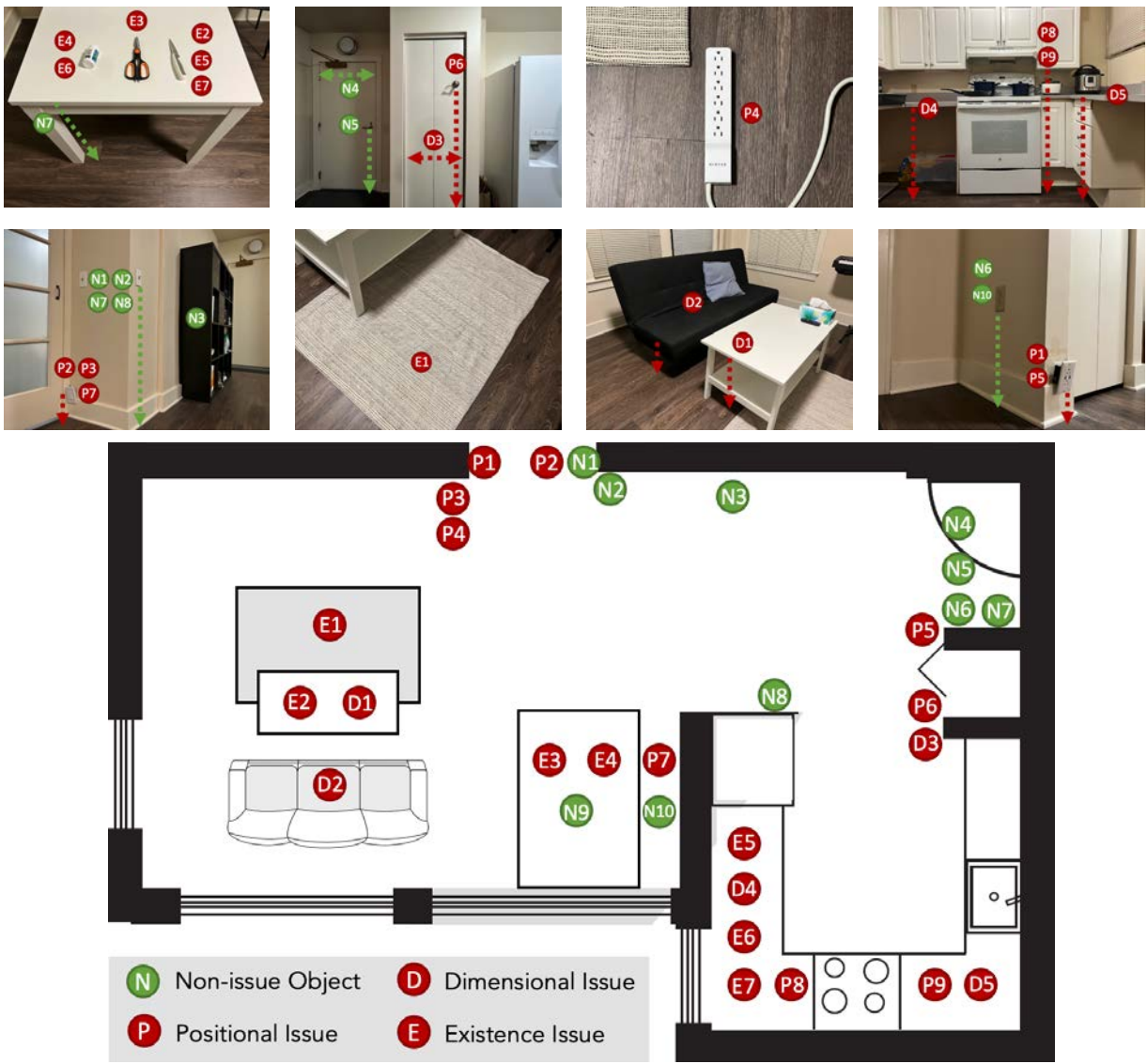

**Figure 5.** The location and example images of the 21 access/safety issues and 10 non-issues (to examine true negatives). Dimensional issues (D) refer to furniture with inaccessible sizes (*e.g.*, a table that is too short or too tall);  Positional issues (P) refer to items placed at an inaccessible height (*e.g.*, a light switch that is out of reach). Existence issues (E) refers to the presence of risky or unsafe items (*e.g.*, a scissors or knife).

**Table 3.** The three scanning factors used in our preliminary evaluation.

| Scanning Factor | Easy | Moderate | Hard |
| --- | --- | --- | --- |
| Lighting Condition | Well lit (> 30 lux) | Partially lit (~5 lux) | Very dark (< 1 lux) |
| Scanning speed | Slow (scan in ≤ 2 mins; sustained hover over objects) | Medium (scan in ≤ 2 mins; sustained hover over objects) | Fast (scan in ≤ 1 min; no sustained hover over objects) |
| Tidiness | Clean | Moderately messy | Very messy |

**Table 4.** Specific measures used to evaluate RASSAR performance.

| Performance Measure | Abbr. | Description |
| --- | --- | --- |
| **True positive** | TP | Successfully identifying an access/safety issue |
| **True negative** | TN | Successfully avoiding a non-issue |
| **False positives** | | |
|     Misclassifications | FP-MISC | Misclassifying an object or reporting the wrong issue for a problem (*e.g.*, classifying a *table* as a *chair*) |
|     Extra detection | FP-E | Classifying a non-issue as a problem |
|     Dimension/position error | FP-DP | Object identified correctly but *dimension* or *position* is incorrect, leading to a false positive (*e.g.*, a table with an accessible height being inferred as too low or high). |
| **False negatives** | | |
|     Missed | FN-M | Missed classifying an object |
|     Dimension/position error | FN-DP | The object has an access/safety issue related to *dimension* or *position* but the system failed to detect it. |

**Table 5.** Aggregate performance measures. FP is a summation of *FP-MISC, FP-E,* and *FP-F* while FN sums *FN-M* and *FN-F*. All measures return a value between 0-1 where 1 is best.

| Measure | Calculation | Description |
| --- | --- | --- |
| **Precision** | $\frac{TP}{TP + FP}$ | The fraction of *true positives* over all made classifications (*i.e.*, when RASSAR makes a classification, how likely is it to be correct?) |
| **Recall** | $\frac{TP}{TP + FN}$ | The fraction of *true positives* over all possible correct classifications (*i.e.*, how good is RASSAR at finding all possible problems?) |
| **F$_1$ Score** | $2 \frac{Precision * Recall}{Precision + Recall}$ | A standard technique to combine both precision and recall together and is generally described as the harmonic mean of the two. |
| **Accuracy** | $\frac{TP + TF}{All\ possible\ cases}$ | Describes the number of correct classifications over all possible classifications (31 in this case). |

Across the six experimental conditions, our accuracy ranged from 48.4-89.7%—see Table 6. Unsurprisingly, the *well lit, clean,* and *medium scan speed* performed best with a precision and recall of 87.4% and 92.8% respectively while the *low light* trials performed worst, primarily affecting recall: *precision*=78.8%; *recall*=26.3%. The second-worst performing trials were the *fast scan speed* condition, which resulted in 91.5% precision, 41.7% recall, and 60% overall accuracy. The amount of clutter did reduce performance from 89.7% to 82.6% (for *messy*) and 72.3% (for *very messy*) but its impact was more limited than lighting and scan speed.

To complement this quantitative analysis, we also provide example identifications and classification errors in Figures 6 and 7. For example, we show true positive examples in Figure 6 (left) as well as examples where RASSAR reported no issues for actual problems (false negatives; Figure 7 right).

**Table 6.** Our evaluation results. Each row specifies three tested factors, which are averaged over 5 scans. TP, FP, TN, and FN are counts. Precision, recall, $F_1$ score, and accuracy are percentages.

| | Factors | | | Ground Truth | | Result Cnts | | | | Result Stats (%) | | | |
|---|---|---|---|---|---|---|---|---|---|---|---|---|---|
| | Light | Tidy | Speed | TP | TN | TP | FP | TN | FN | Prec. | Rec. | $F_1$ | Acc. |
| 1 | Well lit | Clean | Med. | 21 | 10 | 18 | 2.6 | 9.8 | 1.4 | 87.4 | 92.8 | 90 | 89.7 |
| 2 | Partial | Clean | Med. | 21 | 10 | 16.2 | 3.6 | 10 | 2.8 | 81.8 | 85.3 | 83.5 | 84.5 |
| 3 | Poorly lit | Clean | Med. | 21 | 10 | 5.2 | 1.4 | 9.8 | 14.6 | 78.8 | 26.3 | 39.4 | 48.4 |
| 4 | Well lit | Messy | Med. | 21 | 10 | 15.6 | 2.6 | 10 | 4.8 | 85.7 | 76.5 | 80.8 | 82.6 |
| 5 | Well lit | V. messy | Med. | 21 | 10 | 12.4 | 4.4 | 10 | 7.4 | 73.8 | 62.6 | 67.8 | 72.3 |
| 6 | Well lit | Clean | Fast | 21 | 10 | 8.6 | 0.8 | 10 | 12 | 91.5 | 41.7 | 57.3 | 60 |

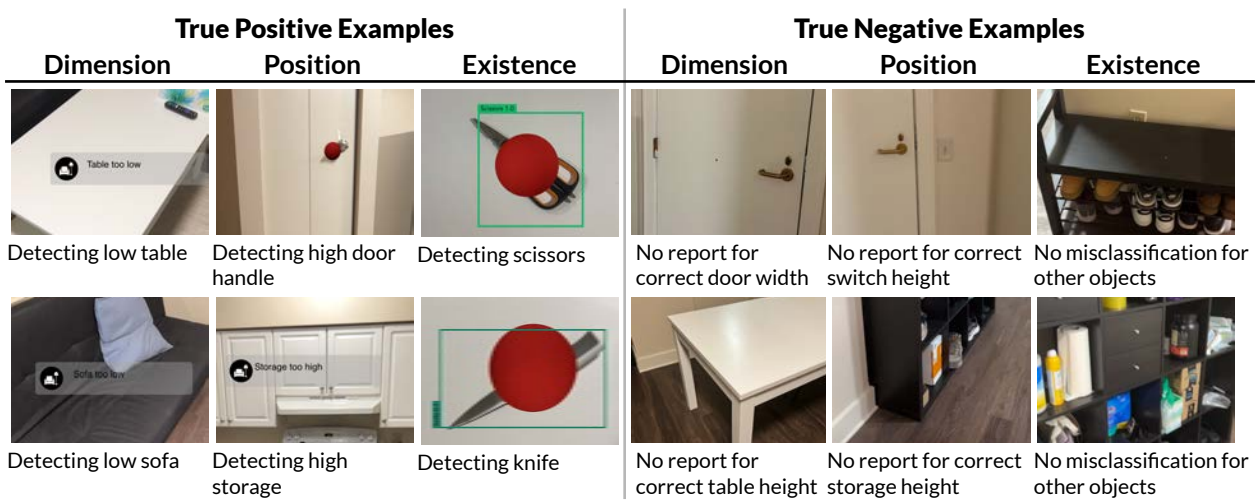

**Figure 6.** Example true positives and true negatives from our evaluation. The red sphere is a preliminary indicator used to indicate the presence of a problem anchored in 3D space and projected in AR.

| False Positive Examples | | False Negative Examples | |
| --- | --- | --- | --- |
| Misclassifications | Extra Detections | Missed Problem | Dimn/Pstn Error |
| 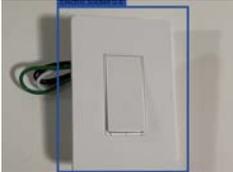 | 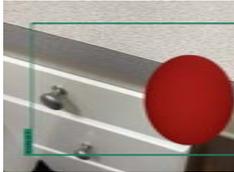 | 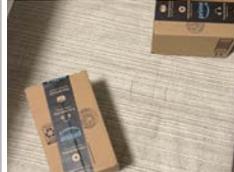 | 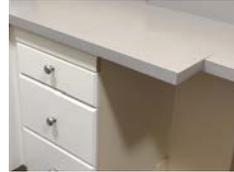 |
| Misclassified switch as socket | Detect counter edge as knife | Failed to detect rug due to clutter on it | Failed to report high counter |
| 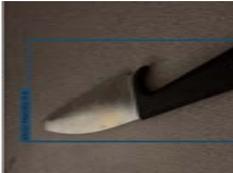 | 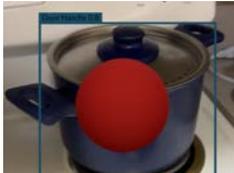 | 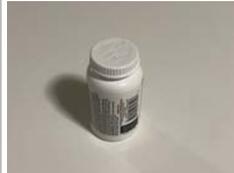 | 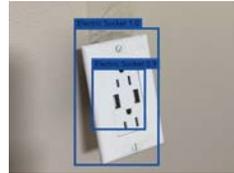 |
| Misclassified knife as door handle | Detect pot as door handle | Failed to detect medication | Failed to report high socket |

**Figure 7.** Example false positive and false negative detections.

# Discussion and Conclusion

In this workshop paper, we introduced RASSAR, an initial prototype for semi-automatically identifying accessibility and safety issues of indoor spaces using LiDAR depth scanning and computer vision. While preliminary, RASSAR demonstrates a new potential approach for mapping and assessing the accessibility and safety of buildings—both public and private. Such a tool could be used to help plan infrastructural renovations (*e.g.*, to add grab bars or ramps) or simply to inform the placement or type of furniture and the reduction of clutter (*e.g.*, electrical cords as tripping hazards) in living spaces (*e.g.*, homes, hotels).

**Study limitations.** Our study helped identify and characterize ideal scanning conditions; however, it was limited both in scope and size. First, only one apartment and a single set of objects were evaluated—future work should examine multiple homes, rooms, and additional object types. Second, a single researcher performed the scan—future studies should examine the effect of different human scanners. Third, and finally, a more thorough evaluation of the custom-trained YOLOv4 detection model should be performed.

**What is semi-automatic?** Recent AI-based companies like *AccessiBe* and *AudioEye* have faced significant criticism for overclaiming how well their tools automatically identify web accessibility problems and provide supposed WCAG-compliant solutions [14,31]. Tools like RASSAR may be subject to similar criticism, albeit for the physical rather than the virtual world. However, our envisioned goal with RASSAR is to create a human+AI system that enables both to work better together. For example, allowing users to actively correct false positives, add new training data for false negatives, and educate them about potential problems. Still, we must confront ethical questions about using imperfect tools to identify problems and must strive for rigorous evaluation and transparency [41]. The ways in which RASSAR actively combines human insights with AI is an active area of future work.

**Future work. Finally, w**ith our initial proof-of-concept prototype created, we would now like to recruit and solicit feedback from key stakeholder groups to refine and update our prototype. We would also like to prototype and examine new functionality for indoor public spaces (*e.g.*, stores, restaurants, government buildings), experiment with AR-based head-mounted displays that could continuously track and update accessibility/safety states by passersby, and examine methods for users to share training data and new JSON specifications. We also plan to improve our deep learning model to improve recognition accuracy and efficiency (*e.g.*, minimize gaze time).

## Acknowledgements

This work is funded in part by the NSF under grant #1652339 and UW CREATE (Center for Research and Education on Accessible Technology and Experiences)

## Author Bios

**Xia Su** is a PhD student in Computer Science at UW interested in creativity support and accessibility. His research explores the intersection of AR, CV, accessibility and design, trying to make indoor spaces a better medium for easier and more creative lives. He has a BS and a MS in Architecture from Tsinghua University.

**Kaiming Cheng** is a PhD student in Computer Science at UW interested in the security and privacy of Augmented Reality/Virtual Reality (AR/VR). More specifically, he is interested in AR/VR systems, identifying the security and privacy flaws, and understanding end-user safety and privacy needs.

**Han Zhang** is a PhD student in Computer Science at UW interested in human behavior modeling and the design of computer-aided methods to improve human well-being. She is also interested in fairness in machine learning (ML) and building explainable ML to better understand human behaviors.

**Jaewook Lee** is an incoming PhD student in Computer Science at UW interested in augmented reality, mobile computing, and accessibility. His research primarily focuses on designing mobile solutions that can facilitate everyday tasks. He is supported by an NSF Graduate Fellowship.

**Yueqian Zhang** is a Master's student in Computer Science at UW with a focus on computer graphics and human-computer interaction. She is interested in human-centered design and developing creative applications and visualizations that spark creativity and empower communities.

**Jon E. Froehlich** is an Associate Professor in Computer Science at UW, the Director of the Makeability Lab, and the Associate Director of the Center for Research and Education on Accessible Technology and Experiences (CREATE). He is interested in applying Human+AI methods to transform how we study and analyze urban accessibility—both indoor and outdoor spaces.

## Rationale for Attendance

Our paper positions the *home* as central to the future of urban accessibility—a place for comfort, for family, and for access and safety. In doing so, our submission moves beyond public infrastructure and transit as core focal points and broadly relates to at least two of the ten key workshop questions: *how is urban accessibility data collected* and *what is the role of AI in assessing accessibility?* Of our six authors, five are students and all are junior scholars who would benefit from the discussions and networking that the workshop would provide—all which could likely shape their dissertation directions.

# Appendix

**Table 7.** A table of the 18 currently supported access/safety issues in RASSAR. RASSAR currently asks the user to select a primary target community but, of course, disabilities and identities can intersect; future versions of the tool will enable multiple selections. Acronyms of literature sources: ADA: ADA Standards for accessible design [52]; HSPD: Home Safety for People With Disabilities [7]; HAC: Home Accessibility Considerations for Your Living Room [30]; FHADM: Fair Housing Act Design Manual [43]; VA: Vision Aware [44]; HSSAT: Home Safety Self-Assessment Tool [33]; Child Safety Checklist [16].

| Issue Category | Object Category | Measurement | Dimension (in inches) | Literature Sources | Older Adults | BLV | Wheelchair Users | Children |
|---|---|---|---|---|---|---|---|---|
| **Object Dimension** | Table | Height | >=28 & <=34 | ADA | | | ✔ | |
| | Counter | Height | >=28 & <=34 | HSPD | | | ✔ | |
| | Toilet | Height | >=17 & <=19 | | | | ✔ | |
| | Sofa | Height | | | | | | |
| | Sink | Height | <=17 | | | | ✔ | |
| | Chair | Depth | | | | | | |
| | Door | Radius | >=32 | ADA | | | ✔ | |
| **Object Position** | Cabinet | Height | <=27 | ADA | | | ✔ | |
| | Knob | Height | >=34 & <=48 | ADA | | | ✔ | |
| | Door Handle | Height | >=34 & <=48 | ADA | | | ✔ | |
| | Light Switch | Height | >=15 & <=48 | FHADM | | | ✔ | |
| | Grab Bar | Height | >=18 & <=27 for child<br>>=33 & <=36 for adults | ADA | ✔ | ✔ | ✔ | ✔ |
| | Electric Socket | Height | >=15 & <=48 | VA, HSSAT, FHADM | ✔ | ✔ | ✔ | ✔ |
| **Object Existence** | Rug | Presence | null | VA, HSSAT | ✔ | ✔ | ✔ | ✔ |
| | Scissors | Presence | null | CSC | | | | ✔ |
| | Knives | Presence | null | CSC | | | | ✔ |
| | Medication | Presence | null | CSC | | | | ✔ |
| | Smoke Alarm | Absence | null | VA, HSPD | ✔ | ✔ | ✔ | ✔ |